\documentclass[aps,prl,twocolumn,superscriptaddress,amsmath]{revtex4-1}

\usepackage{amsmath}
\usepackage{amssymb}
\usepackage{amsthm}
\usepackage{bbm}
\usepackage{enumerate}
\usepackage{color}
\usepackage{graphicx}
\usepackage{tikz}

\usetikzlibrary{backgrounds,fit,decorations.pathreplacing}  

\usetikzlibrary{decorations.pathmorphing}
\usetikzlibrary{positioning}
\usetikzlibrary{shapes.callouts}

\theoremstyle{plain}
\theoremstyle{definition}

\newcommand{\tr}{\operatorname{tr}}
\newcommand{\ket}[1]{\ensuremath{\vert#1\rangle}}
\newcommand{\bra}[1]{\ensuremath{\langle #1|}}

\usepackage[normalem]{ulem}
\newcommand{\ens}[0]{\ensuremath}
\newcommand{\x}[0]{\ens{\otimes}} 
\renewcommand{\phi}[0]{\ens{\varphi}} 
\newcommand{\iE}[0]{\ens{\mathrm{i}}}
\newcommand{\EZ}[0]{\ens{\mathrm{e}}} 
\newcommand{\Eins}[0]{\ens{\mathbbm{1}}}
\newcommand{\diag}[0]{\ens{\mathrm{diag}}}
\newcommand{\M}[0]{\ens{\mathcal{M}}}
\newcommand{\F}[0]{\ens{\mathbb{F}}}
\newcommand{\N}[0]{\ens{\mathbb{N}}}

\newcommand{\C}[0]{\ens{\mathbb{C}}}

\newcommand{\Mg}[1]{\ens{\left\lbrace #1 \right\rbrace}}

\newcommand{\MgE}[1]{\ens{\Mg{1,\dots,#1}}}

\begin{document}

\title{A new class of mutually unbiased bases for state estimation in qubit registers}

\author{Ulrich Seyfarth}
\affiliation{Institut f\"ur Angewandte Physik, Technische Universit\"at Darmstadt, Hochschulstra\ss{}e 4a, 64289 Darmstadt, Germany}

\author{Kedar S. Ranade}
\affiliation{Institut f\"ur Quantenphysik, Universit\"at Ulm, Albert-Einstein-Allee 11, 89081 Ulm, Germany}

\date{October 28, 2011}


\begin{abstract}
  For the complete estimation of arbitrary unknown quantum states by measurements, the use of mutually unbiased
  bases has been well-established in theory and experiment for the past 20 years. However, most constructions of these
  bases make heavy use of abstract algebra and the mathematical theory of finite rings and fields, and no simple
  and generally accessible construction is available. This is particularly true in the case of a system composed
  of several qubits, which is arguably the most important case in quantum information science and quantum computation.
  In this letter, we close this gap by providing a simple and straightforward method for the construction of
  mutually unbiased bases in the case of a qubit register. We show that our construction is also accessible to
  experiments, since only Hadamard and controlled phase gates are needed, which are available in most practical
  realizations of a quantum computer. Moreover, our scheme possesses the optimal scaling possible, i.\,e.,
  the number of gates scales only linearly in the number of qubits.
\end{abstract}

\maketitle

In quantum physics, the estimation of the state of a system is of high practical value \cite{PR04}. It is well-known that
for the complete estimation of a state, known as state tomography, a single measurement is not sufficient, even if
performed many times to get the statistics of such measurement. It is necessary to measure a state in various
different bases. The best choice of such bases for an arbitrary system is so-called mutually unbiased bases
(MUBs), which offer the highest information outcome, as already stated by Wootters and Fields \cite{WoottersFields}:
\textit{mutually unbiased bases provide an optimal means of determining an ensemble's state}. Experimental results
demonstrate the practicability of those schemes \cite{AdamsonSteinberg,Nagali_etal,Lima}.
Different construction methods for MUBs are known \cite{Schwinger,Alltop,Ivanovic,WoottersFields,Bandopadyay-ua,KlappeneckerRoetteler}.
For a $d$-level system, i.\,e., a system described by a $d \times d$ density matrix, one would need $d+1$
mutually unbiased bases, since any measurement statistically reveals $d-1$ parameters. Unfortunately, it is
not even known whether $d+1$ such bases exist in every $d$-level system. Mutually unbiased bases are related
to different topics in mathematics and physics, e.\,g., quantum cryptography, foundations of physics \cite{MaassenUffink}, orthogonal Latin squares or hidden-variable models \cite{Paterek-ua-1,Paterek-ua-2} and even Feynman's path integral \cite{Tolar}.
\par In this letter, we want to focus on a system which is of particular interest in quantum information processing,
namely, a quantum register built of qubits. We propose a complete set of mutually unbiased bases for quantum
registers of size $1, 2, 4, 8, \ldots, 256$. In general, the construction of MUBs is quite involved and uses
methods from abstract algebra and the mathematical theory of finite fields and rings \cite{KlappeneckerRoetteler}, which are far apart from most
methods that are commonly used in physics.
We overcome this problem in such a way, that our construction (although not its proof) is very easy to follow
and to apply by anyone with just basic knowledge of linear algebra. Moreover, our construction is applicable to
experiments with only limited effort. In particular, the experimentator must only be able to perform, on a system
of $m$ qubits, a single unitary operation $U_m$, by which all the $d+1$ MUBs are generated. (Obviously, he must be
able to perform measurements in at least one basis, say, the standard basis.)
This single unitary operator is of the particular form $U_m = \EZ^{\iE\psi_m} 2^{-m/2} H^{\x m} P_m$, where
$2^{-m/2}H^{\x m}$ denotes the $m$-fold tensor product of Hadamard matrices, $P_m$ is a diagonal phase matrix and
$\EZ^{\iE\psi}$ a global phase. The columns of $U,\,U^2,\,U^3,\,\dots,\,U^{2^m+1} = \Eins_{2^m}$ then define mutually
unbiased bases.
Any phase matrix $P_m$ can be decomposed into \texttt{CPhase} gates and together with the one-qubit Hadamard gate,
belongs to the building blocks of one of the several universal sets of gates for a quantum computer. As any
reasonable quantum computer will be able to perform such operations \cite{DiVincenzo}, no extra effort is needed to
implement our measurement; an example circuit is given in the figure.

\par We shall now present our construction in detail. For a single qubit we have the Hadamard transformation
$2^{-1/2} H$ with $H = {\tiny \begin{pmatrix} 1 & 1 \\ 1 & -1 \end{pmatrix}}$, which switches between the
mutually unbiased $z$- and the $x$-bases.  (It is convenient to choose the matrix $H$ to be non-normalized,
so that it and its tensor products contain only $\pm 1$.)
Multiplying this by a phase $P_1 = \diag(1,-\iE)$, we find that $T := 2^{-1/2} H \cdot P_1$ cyclically
switches between the mutually unbiased $z$-, $x$- and the $y$-bases (up to a global phase). This matrix is well-known
e.\,g., in quantum key distribution where it is used in the six-state protocol (cf. e.\,g. \cite{Gottesman,GottesmanLo}).
We want to generalize this
construction to higher numbers of qubits. If we take a number $m$ of qubits, it is obvious that the $m$-fold tensor
product $H^{\x m}$ switches between two mutually unbiased bases, composed of local $z$- and $x$-bases. We can
switch between $2^m+1$ mutually unbiased bases, if we apply a diagonal phase gate (such as $P_1$ above). Determining
these local phases, which turn out to be either $\pm 1$ or $\pm \iE$ is non-trivial, not even numerically, but
can be achieved by our method. Given the unitary $T$ as above, we can construct a two-qubit phase matrix by reading
it out row-wise, i.\,e. $P_2 = \diag(1,-\iE,1,\iE)$. It turns out that $U_2 = \iE\,2^{-1}H^{\x 2}P_2$ produces a
cycle of five MUBs. We can iterate this procedure to get MUBs for 4,\,8,\, etc. qubits. Although these choices of
$P_4,\,P_8,$
etc. appear to be accidental for now, this structure has deep mathematical roots, which we will elaborate
on later in this letter. In particular, this construction is related to Wiedemann's conjecture \cite{W88} from
finite-field theory, and is, as such, only valid for up to 256 qubits. However, it may hold for even higher
numbers such as 512, 1024 etc. If this is true for all powers of two, this would prove Wiedemann's conjecture.
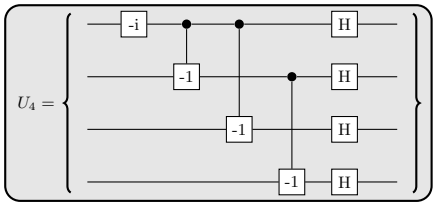
\begin{figure}[t]
  \centerline{
    \begin{tikzpicture}[scale=0.7, transform shape] 
    %
    \tikzstyle{operator} = [draw,fill=white,minimum size=1.5em] 
    \tikzstyle{phase} = [fill,shape=circle,minimum size=5pt,inner sep=0pt]
    \tikzstyle{surround} = [fill=black!10,thick,draw=black,rounded corners=2mm]
    %
    \node at (1,0) (q1) {};
    \node at (1,-1) (q2) {};
    \node at (1,-2) (q3) {};
    \node at (1,-3) (q4) {};
    %
    \node[operator] (op11) at (2,0) {-i} edge [-] (q1);
    \node[phase] (op12) at (3,0) {} edge [-] (op11);
    \node[operator] (op21) at (3,-1) {-1} edge [-] (q2);
    \draw[-] (op12) -- (op21);
    %
    \node[phase] (op13) at (4,0) {} edge [-] (op12);
    \node[operator] (op31) at (4,-2) {-1} edge [-] (q3);
    \draw[-] (op13) -- (op31);
    \node[phase] (op22) at (5,-1) {} edge [-] (op21);
    \node[operator] (op41) at (5,-3) {-1} edge [-] (q4);
    \draw[-] (op22) -- (op41);
    \node[operator] (op14) at (6,0) {H} edge [-] (op13);
    \node[operator] (op23) at (6,-1) {H} edge [-] (op22);
    \node[operator] (op32) at (6,-2) {H} edge [-] (op31);
    \node[operator] (op42) at (6,-3) {H} edge [-] (op41);
    \draw[-] (op14) -- (7,0);
    \draw[-] (op23) -- (7,-1);
    \draw[-] (op32) -- (7,-2);
    \draw[-] (op42) -- (7,-3);
    %
    \draw[decorate,decoration={brace},thick] (7.3,0.2) to
	node[midway,right] (bracketR) {}
	(7.3,-3.2);
    \draw[decorate,decoration={brace},thick, rotate=180] (-0.8,3.2) to
	node[midway,left,rotate=180] (bracketL) {$U_4 =\;\;$}
	(-0.8,-.2);
    %
    \begin{pgfonlayer}{background} 
    \node[surround] (background) [fit = (op11) (op42) (bracketL) (bracketR)] {};
    \end{pgfonlayer}
    \end{tikzpicture}
  }

  \caption{Quantum circuit for the generator $U_4$ for a four-qubit register.}
\end{figure}
\par To implement these unitaries in experiments we will now give decomposition of the phase system $P_m$ into
elementary gates. To do so, we will have to touch the mathematical roots of this matrix structure. But let us
start with two elementary gates, the one-qubit \texttt{Phase} gate and the two-qubit controlled-phase or
\texttt{CPhase} gate:
\begin{align}
  \texttt{Phase}_q(\EZ^{\iE\phi}) &= \ket{0}_q\bra{0} + \EZ^{\iE \phi} \ket{1}_q\bra{1},\\
  \texttt{CPhase}_{s \rightarrow t}(\EZ^{\iE\phi}) &= \ket{0}_s\bra{0} \x \ket{0}_t\bra{0}
    + \ket{0}_s\bra{0} \x \ket{1}_t\bra{1} \nonumber\\
    &\hspace{-1cm}+ \ket{1}_s\bra{1} \x \ket{0}_t\bra{0}
    + \EZ^{\iE\phi} \ket{1}_s\bra{1} \x \ket{1}_t\bra{1}.
\end{align}
The prototype of a phase gate is Pauli's $\sigma_z$ with $\EZ^{\iE\phi} = -1$, but in our case we usually
have $\EZ^{\iE\phi} = \pm \iE$. Note that in the case of a \texttt{CPhase} gate, source and target qubits are
interchangeable and moreover, all these gates commute, since they are diagonal.
\par The recursive procedure we use to construct $P_{2m}$ from $P_m$ results from a similar procedure for an
$m \times m$ matrix $B_m = (b_{kl})_{k,l = 1}^{m}$ with entries $0$ and $1$ and arithmetics modulo $2$ (more
formally, over the finite field $\F_2$), which describes the phase system. This matrix has to fulfill several
properties \cite{KRS10,SR11} related to a stabilizer structure (that is, why we call it the reduced stabilizer
matrix), e.\,g. it must be symmetric. However, in our case it is more advisable to view it as the adjacency matrix
of an undirected graph with $m$ qubits (as vertices) and edges between qubits $i$ and $j$, if $b_{ij} = 1$.
We allow loops, i.\,e. $b_{ii} = 1$ is possible, but they need to be treated separately.
\par We shall label the basis vectors of an $m$-qubit system by bit strings of length $m$, namely by
$\vec{\j} = (j_1,\,\dots,\,j_m)^t$, where $j_i \in \Mg{0,\,1}$ for all $i \in \MgE{m}$. The relation between
$B_m$ and the phase system $P_m = \diag\bigl[(p_{\vec{\j}})_{\vec{\j} \in \F_2^m}\bigr]$ was derived before \cite{KRS10,SR11} and can be written as
\begin{equation}\label{eqn:pj_Gatter}
  p_{\vec{\j}} = \iE^{\sum_{k,l = 1}^{m} b_{kl} j_k j_l} (-1)^{\sum_{k = 1}^{m} b_{kk}j_k}.
\end{equation}
Note that the sums are \emph{not} performed modulo 2. We may rewrite the first factor as $\prod_{k,l = 1}^{m}
(\iE^{j_k j_l})^{b_{kl}}$ or $\prod_{k < l} ((-1)^{j_k j_l})^{b_{kl}} \cdot \prod_{k}
(\iE^{j_k})^{b_{kk}}$, as $B_m$ is symmetric. The second factor of this expression may be joined with the second
factor of $p_{\vec{\j}}$ to find
\begin{equation}\label{eqn:pj_product}
  p_{\vec{\j}} = \prod\nolimits_{k,l=1,k<l}^{m} ((-1)^{j_k j_l})^{b_{kl}} \cdot \prod\nolimits_{k=1}^m ((-\iE)^{j_k})^{b_{kk}}.
\end{equation}
This can be interpreted as follows:
\begin{itemize}
  \item If $b_{kk} = 1$, perform $\texttt{Phase}_k(-\iE)$.
  \item If $b_{kl} = 1$ for $k<l$, perform $\texttt{CPhase}_{k \rightarrow l}(-1)$.
\end{itemize}
The experimental resources (together with the Hadamard gate) are therefore precisely the same as for the
preparation of graph states. More precisely, the number of gates is directly related to the number of non-zero entries
in $B$.
\par We go on to discuss the doubling scheme from $m$ to $2m$ qubits explicitly. We start with a single qubit
and $B_1 = (1)$. To go from $m$ qubits to $2m$ qubits, we use the block-matrix mapping
\begin{equation}\label{eqn:B2m}
  B_m \mapsto B_{2m} = \begin{pmatrix} B_m & \Eins_m \\ \Eins_m & 0_m \end{pmatrix}.
\end{equation}
For example, for one to four qubits, the mapping reads
\begin{equation}
  (1) \mapsto \begin{pmatrix} 1 & 1 \\ 1 & 0 \end{pmatrix} \mapsto
    \begin{pmatrix} 1 & 1 & 1 & 0 \\ 1 & 0 & 0 & 1 \\ 1 & 0 & 0 & 0 \\ 0 & 1 & 0 & 0\end{pmatrix} \quad\text{etc.}
\end{equation}
To check that these matrices $B_m$ fulfill the criteria for generating mutually unbiased bases \cite{KRS10} is
in general non-trivial, but is in our case for $m \in \Mg{1,\,2,\,4,\,8,\,\dots,\,256}$ guaranteed by 
previous results \cite{SR11} in connection with the work of Wiedemann \cite{W88}.
\par Note that the number of non-zero entries scales linearly in $m$. The only diagonal non-zero element
is $b_{11}$ for any number $m$, so there is just a single \texttt{Phase} gate, and eq.~(\ref{eqn:pj_product}) implies that
the value of $p_{\vec{\j}}$ with $\vec{\j} = (j_1,\,\dots,\,j_m)^t$ is real, if $j_1 = 0$ and purely imaginary otherwise.
The number of \texttt{CPhase} gates, which is the number of upper diagonal elements, increases by $m$, if we go from
$m$ to $2m$ qubits. Altogether we have a single \texttt{Phase} gate and $m-1$ \texttt{CPhase} gates for a system
composed of $m$ qubits, i.\,e. linear scaling with the number of qubits.
\par By now, we have constructed a matrix which cyclically switches between $2^m+1$ mutually unbiased bases in two
ways: first, by using the explicit doubling scheme for a unitary phase matrix $P_m$ and second, by invoking a
matrix $B_m$ for a decomposition into elementary gates.
Although not directly relevant to experiments, the matrix $V_m = H^{\x m}P_m$ still lacks normalization and
a global phase $\EZ^{\iE\psi_m}$ to give the generating matrix $U_m$, such that $U_m^{2^m+1} = \Eins$.
It is necessary that the
$2^m$ eigenvalues of $U_m$ are ($2^m+1$)-th roots of unity. Using a conjecture, which was numerically checked,
that $U_m$ and therefore $V_m$ are non-degenerate \cite{KRS10}, we may choose $\lambda = 1$ not to be an eigenvalue
of $U_m$. As the characteristic polynomial of $U_m$ is then given by $\chi(\lambda) = \det(\lambda \Eins_{2^m} - U_m)
= (\lambda^{2^m+1}-1)/(\lambda-1) = \sum_{k = 0}^{2^m} \lambda^k$, we see that we have to divide our matrix $V_m$
by $-\tr V_m$ to get $U_m$. This trace is given by $1+\iE$ for $m = 1$ and $\iE 2^{m/2}$ for $m \in \{2,4,8,\ldots, 256\}$, which
we derive later in this letter.
\par For a single-qubit register we take the solution that we find with the help of matrix $B_1 = (1)$, i.\,e.
$V_1 = \tiny H \cdot \diag(1,-\iE)$, where $\diag(\ldots)$ maps values from a $d$~dimensional row vector to the
diagonal of a $d \times d$~matrix. By dividing $V_1$ by $-\tr V_1 = -(1+\iE)$ we receive the matrix $U_1$ that switches between three mutually unbiased bases of a single qubit. To get an operator that switches between five mutually
unbiased bases for a two-qubit register, we introduce a \emph{chop-map} $\M$ that behaves in the following way:
For $V$ being an arbitrary $d \times d$ matrix, $v = \M(V)$ is a $d^2$-dimensional row vector, where
the first row of $V$ is mapped to $v_1,\,\ldots,\,v_d$, the second row of $V$ is mapped to $v_{d+1},\,\ldots,\,v_{2d}$
etc. If we use this vector as a phase vector such as $V_2 = \tiny H^{\otimes 2} \cdot \diag( \M(V_1) )$ and apply
normalization and the associated global phase, i.\,e. $U_2 = V_2 / (-\tr V_2)$, we obtain the desired operator for
a two-qubit register. As stated before, this construction works until a register length of $256$, and it holds
for an arbitrary number of register length doublings, if and only if Wiedemann's conjecture is correct.
\par In the following we will show how to transfer the doubling scheme of $B_m$ as in eq. \eqref{eqn:B2m}
to the doubling scheme of $V_m$ and $U_m$. We need a particular ordering of the entries of the basis vectors
in $U_m$. We label each vector with an $m$-bit string $\vec{\i} = (i_1,\,\dots,\,i_m)^t$ of zeros and ones and sort
them in binary increasing order. For the non-normalized Hadamard matrix this results in Sylvester's construction
$H_{m} \in M_{2^m}(\C)$ for $m \in \N_0$. For $m = 0$, we have $H_0 := (1)$ and recursively define
\begin{equation}
  H_{m+1} := \begin{pmatrix}H_{m}&H_{m}\\H_{m}&-H_{m}\end{pmatrix} \in M_{2^{m+1}}(\C);
\end{equation}
we see that $H = H_1$ is the well-known regular Hadamard matrix in the qubit case. It is obvious that the entries
in every $H_{m}$ are either $+1$ or $-1$ and further, $H_{m}$ can be seen as the explicit construction of the
$m$-fold tensor product $H^{\x m}$.

\par We can now relate the chop-map doubling scheme for $P_m$ to the doubling scheme for $B_m$ of eq.~\eqref{eqn:B2m}.
We want to construct the explicit phases $p^\prime_{\vec{\j}^\prime}$ of $P_{2m}$ to the phases $p_{\vec{\j}}$ of
$P_m$. As the length of $\vec{\j}^\prime$ is twice the length of $\vec{\j}$, we may write
$\vec{\j}^\prime = (\vec{\j}_1,\,\vec{\j}_2)^t$. In our case of $B_m$, the exponent of the second factor of
$p_{\vec{\j}}$ as in eq.~(\ref{eqn:pj_Gatter}) depends only on $b_{11}$ and remains unchanged by the doubling. The exponent of the first
factor can be separated into two parts, one arising from the old $B_m$ and another arising from the doubling.
By invoking eq.~\eqref{eqn:pj_Gatter} this directly results in the exponent being $\vec{\j}_1 \cdot B_m \vec{\j}_1 + 2 \vec{\j}_1 \cdot \vec{\j}_2$.
The first part is exactly the phase system of the old matrix $P_m$, while working out the second part results in
a factor $(-1)^{\vec{\j}_1 \cdot \vec{\j}_2}$, which represents the old non-normalized Hadamard matrix, where
$\vec{\j}_1$ and $\vec{\j}_2$ indicate rows and columns, respectively.
We conclude that the calculation of the phases of $V_{2m}$ can be done by applying the local phases of $V_m$ to a Hadamard matrix of the right size and then concatenating the resulting rows one after the other.

\par Before concluding this letter, we shall provide the reader with an explicit form of the matrix $U_m$, which
generates the complete set of mutually unbiased bases; to this end, we derive the form of $V_m$ and then calculate
its trace, so that $U_m = -V_m / \tr V_m$. As seen above, we may write the matrices $V_m$ recursively as
$V_{2m} = H^{\otimes 2m} \mathrm{diag}(\M(V_m))$ with $V_1 = \tiny \begin{pmatrix} 1 & - \iE \\ 1 & \iE \end{pmatrix}
= H \begin{pmatrix} 1 & 0 \\ 0 & -\iE \end{pmatrix}$.
Since the entries of $H$ are given by $(-1)^{\vec{\i} \cdot \vec{\j}}$, we can write $V_1$ as:
\begin{align}
  (V_1)_{i,j} = 
  \begin{cases}
    1 \cdot (-1)^{\vec{\i} \cdot \vec{\j}},    & \text{if  } j \equiv 0 \mod 2,\\
    -\iE \cdot (-1)^{\vec{\i} \cdot \vec{\j}}, & \text{if  } j \equiv 1 \mod 2,
  \end{cases}
\end{align}
with $\vec{\i}$ representing the rows and $\vec{\j}$ the columns of $V_1$ and the indices starting with zero. In the $V_1$ case, $\vec{\i}$ and $\vec{\j}$ are one bit long, respectively. In the $V_2$ case they are two bits long, for the $V_4$ case four bit and so forth, so we write the vector as a bit string with the lowest bit on the left like
$\vec{\i} = (i_1,\,\ldots,\,i_m)^t$. To calculate $V_2$, we have to chop-map $V_1$ to a diagonal matrix and multiply
this to $H^{\otimes 2}$. Thus we get for $V_2$:
\begin{align*}
  (V_2)_{i,j} = 
  \begin{cases}
    1 \cdot (-1)^{\vec{\i} \cdot \vec{\j}} \cdot (-1)^{j_1 \cdot j_2},    & \text{if  } j \equiv 0 \mod 2,\\
    -\iE \cdot (-1)^{\vec{\i} \cdot \vec{\j}} \cdot (-1)^{j_1 \cdot j_2}, & \text{if  } j \equiv 1 \mod 2.
  \end{cases}
\end{align*}
By iteration, this results in the expression
\begin{align}\label{eqn:Vk}
  (V_m)_{i,j} =
  \begin{cases}
    1 \cdot (-1)^x,    & \text{if  } j \equiv 0 \mod 2,\\
    -\iE \cdot (-1)^x, & \text{if  } j \equiv 1 \mod 2,\\
  \end{cases}
\end{align}
with $x = (i_1,\,\ldots,\,i_m)^t \cdot (j_1,\,\ldots,\,j_m)^t + (j_1,\,\ldots,\,j_{\frac{m}{2}})^t
\cdot (j_{\frac{m}{2}+1},\,\ldots,\,j_m)^t + \ldots + j_1 \cdot j_2$.
We shall now derive the trace of $V_m$. To simplify matters we consider the real and imaginary
part of $V_m$ separately. Since the trace is the sum over the diagonal entries, we set $\vec{\i} = \vec{\j}$ in
eq. (\ref{eqn:Vk}).
\par For the real part, $j_1=0$ holds due to eq. \eqref{eqn:Vk}. To calculate the sum of the real part, we add the terms $\vec{\j}$ for which $j_{\frac{m}{2}+1} = 0$ pairwise to those terms $\vec{\j}^{\prime}$ where $j^{\prime}_{\frac{m}{2}+1} = 1$. 
The Hamming weight of $\vec{\j}$ and $\vec{\j}^{\prime}$ differs by one, so $(-1)^{\vec{\j} \cdot \vec{\j}} + (-1)^{\vec{\j}' \cdot \vec{\j}'} = 0$.
All subsequent terms of $\vec{\j}$ and $\vec{\j}^{\prime}$ are equal, since $\vec{\j}$ and $\vec{\j}^{\prime}$ differ only in the right half of their bits. The first terms are equal due to the fact, that the lowest bits of $\vec{\j}$ and $\vec{\j}^{\prime}$ are zero. Thus the real part of $\tr V_m$ vanishes. For $m=1$ the real part equals one, since there is only a single term with $j_1=0$.\par
The imaginary part of $\tr V_m$ is given by the sum over every second element of the diagonal, i.\,e. $j_1=1$. We will split this calculation into two steps. In the first step, we take those terms
where the left half of $\vec{\j}$ is filled by ones. One of those terms is given by
\begin{align}
 -(-1)^{\vec{\j} \cdot \vec{\j} + (1\ldots 1) \cdot (j_{\frac{m}{2}+1} \ldots j_m) + (1\ldots 1) \cdot (1\ldots 1) + \ldots + 1 \cdot 1}.
\end{align}
The first summand in the exponent calculates the Hamming weight of $\vec{\j}$, whereas the second summand calculates the Hamming weight of the right half of $\vec{\j}$. Since $m$ is a power of two, for those cases with $m>1$, the first two summands give the same result modulo two. All subsequent summands without the last one result in powers of two, the last term, $1\cdot1$, gives one. For $m=1$ we have only this last summand, so in all cases of $m$ the summands add up to an odd integer, thus these terms give one. There are $2^{m/2}$  ways for $\vec{\j}$ to have only ones in the left half, so they contribute with $2^{m/2}$ to the imaginary part of the trace.
\par There remain those elements of $\vec{\j}$ for which the left half has at least one zero bit, but the lowest bit has to be one for the imaginary elements. We can pair them like in the real part and the same argument brings their sum to zero. Thus the trace of $V_m$ is given by $\iE 2^{m/2}$, and we find
\begin{align*}
  U_m = \iE 2^{-m/2} H^{\otimes m} \diag(\M(  H^{\otimes m/2} \cdot \ldots \cdot H \diag(1,-\iE) ))
\end{align*}
for $m \in \Mg{2,4,8,\ldots,256}$; in the case $m = 1$ we have
$U_1 = -H \diag(1,-\iE)/(1+\iE)$.
\par To summarize, we have shown in this letter how to construct a maximal set of mutually unbiased bases for a
quantum system composed of qubits by a single unitary generator, and we have shown how this operation can
be decomposed into Hadamard, phase and controlled phase gates. The necessary resources to implement our scheme
may be compared to those for preparing graph states, and the number of gates scales only linearly in the number
of qubits. This scaling is optimal, since the graph must be connected. We believe that our approach may be of
interest in the one-way quantum computer by Rau{\ss}endorf and Briegel \cite{BriegelRaussendorf,RaussendorfBriegel,
RaussendorfBrowneBriegel,Walther}, where one uses a nearest-neighbor Ising-type interaction to generate a cluster
state and one-qubit measurements. In our case, we only need measurements in the standard bases, but due to our
construction, we may need \texttt{CPhase} gates with possibly long distances between source and target qubits. It would be useful to overcome
this limitation by a new construction which uses band matrices of limited size, or to use qubit implementations
which make such long-distance \texttt{CPhase} gates possible in an experiment.

\par A slight disadvantage of our system is that we restrict ourselves to numbers of qubits which are powers
of two. Further work may continue in finding generators of mutually unbiased bases for qubit registers of different length.
For example, if $m = 3$, we can choose any of the matrices $B_m$ with $m=2^k$.
A doubling scheme similar to eq. (\ref{eqn:B2m}), more precisely
\begin{equation}
  B_m \mapsto B_{3m} = \begin{pmatrix} B_m & B_m & B_m \\ B_m & B_m & 0_m\\ B_m & 0_m & 0_m \end{pmatrix},
\end{equation}
produces mutually unbiased bases at least for $m \in \Mg{6,\,12,\,24}$.
But investigations to be performed in more detail require a more thorough understanding of the mathematical
principles underlying this construction, but this is beyond the scope of this letter.

\par The authors acknowledge financial support by CASED and BMBF/QuOReP.


\begin{thebibliography}{99}
  \bibitem{PR04}%
    M. Paris and J. \v{R}eh\'a\v{c}ek (eds.), \textit{Quantum state estimation} (Springer-Verlag, 2004)
  \bibitem{WoottersFields}
    W.~K. Wootters and B.~D. Fields, Annals of Physics \textbf{191} (1989), 363--381
  \bibitem{AdamsonSteinberg}
    R.~B.~A. Adamson and A.~M. Steinberg, Phys. Rev. Lett. \textbf{105} (2010), 030406
  \bibitem{Nagali_etal}
    E.~Nagali, L.~Sansoni, L.~Marrucci, E.~Santamato and F.~Sciarrino,
    Phys. Rev. A \textbf{81} (2010), 052317
  \bibitem{Lima}
    G. Lima, L. Neves, R. Guzm\'an, E.~S. G\'omez, W.~A.~T. Nogueira, A. Delgado, A. Vargas, and C. Saavedra, Opt. Express \textbf{19} (2011), 3542--3552
  \bibitem{Schwinger}
    J. Schwinger, Proc. Nat. Acad. Sci. \textbf{46} (1960), 570--579
  \bibitem{Alltop}
    W.~O. Alltop, IEEE Trans. Inf. Theory \textbf{26} (1980), 350--354
  \bibitem{Ivanovic}
    I.~D. Ivanovi\'{c}, J. Phys. A \textbf{14} (1981), 3241--3245
  \bibitem{Bandopadyay-ua}
    S. Bandyopadhyay, P. O. Boykin, V. Roychowdhury and F. Vatan, Algorithmica \textbf{34} (2002), 512--528;
    arXiv:quant-ph/0103162
  \bibitem{KlappeneckerRoetteler}
    A. Klappenecker and M. R{\"o}tteler, Lecture Notes in Computer Science, \textbf{2948} (2004), 137--144
  \bibitem{MaassenUffink}
    H. Maassen, J. B. M. Uffink, Phys. Rev. Lett. \textbf{60} (1988), 1103--1106 
  \bibitem{Paterek-ua-1}
    T. Paterek, M. Paw\l{}owski, M. Grassl and \v{C}. Brukner, Phys. Scr., \textbf{T140} (2010), 014031
  \bibitem{Paterek-ua-2}
    T. Paterek, B. Daki\'c and \v{C}. Brukner, Phys. Rev. A, \textbf{79} (2009), 012109
  \bibitem{Tolar}
    J. Tolar and G. Chadzitaskos, J. Phys. A, \textbf{42} (2009), 245306
  \bibitem{DiVincenzo}
    D. P. DiVincenzo, Fortschritte der Physik \textbf{48} (2000), 771; arXiv:quant-ph/0002077v3
  \bibitem{Gottesman}
    D. Gottesman, Phys. Rev. A \textbf{57} (1998), 127--137
  \bibitem{GottesmanLo}
    D. Gottesman and H.-K. Lo, IEEE Trans. Inf. Theory \textbf{49} (2003), 457--475
  \bibitem{W88}%
    D. Wiedemann, Fib. Quart. \textbf{26} (1988), 290--295
  \bibitem{KRS10}
    O. Kern, K. S. Ranade and U. Seyfarth, J. Phys. A \textbf{43} (2010), 275305; arXiv:quantph/09124661v1
  \bibitem{SR11}
    U. Seyfarth and K. S. Ranade, arXiv:1104.0202v1
  \bibitem{BriegelRaussendorf}
    H. J. Briegel and R. Raussendorf, Phys. Rev. Lett. \textbf{86} (2001), 910--913
  \bibitem{RaussendorfBriegel}
    R. Raussendorf and H. J. Briegel, Phys. Rev. Lett. \textbf{86} (2001), 5188--5191    
  \bibitem{RaussendorfBrowneBriegel}
    R. Raussendorf, D. E. Browne and H. J. Briegel, Phys. Rev. A \textbf{68} (2003), 022312
  \bibitem{Walther}
    P. Walther \textit{et al.}, Nature \textbf{434} (2005), 169--176
\end{thebibliography}
\end{document}